\documentclass{INTERSPEECH2023}
\interspeechcameraready

\usepackage[capitalise, nameinlink, noabbrev]{cleveref}
\usepackage[acronym, shortcuts, nohypertypes={acronym}]{glossaries}
\usepackage{multibib}
\usepackage{graphicx}
\usepackage{tabularx}
\usepackage{listings}

\usepackage{epstopdf}
\usepackage[utf8]{inputenc}

\usepackage{hyperref}
\usepackage{xstring}

\usepackage{color}

\usepackage[utf8]{inputenc}
\usepackage{graphicx}
\graphicspath{ {./images/} }

\title{Happy or Evil Laughter? Analysing a Database of Natural Audio Samples}
\name{Aljoscha Düsterhöft\,$^{1}$, Felix Burkhardt$^{2}$, Bj\"orn W. Schuller\,$^{2,3,4}$}
\address{$^1$Technical University of Berlin, Germany, $^1$audEERING GmbH, Germany, \\$^3$Chair EIHW, University of Augsburg, Germany, $^4$GLAM, Imperial College London, UK}
\email{alj.duesterhoeft@gmail.com, [fburkhardt, bs]@audeering.com}

\begin{document}

\maketitle

\begin{abstract}
We conducted a data collection on the basis of the Google AudioSet database by selecting a subset of the samples annotated with \textit{laughter}. The selection criterion was to be present a communicative act with clear connotation of being either positive (laughing with) or negative (being laughed at).
On the basis of this annotated data, we performed two experiments: on the one hand, we manually extract and analyze phonetic features. On the other hand, we conduct several machine learning experiments by systematically combining several automatically extracted acoustic feature sets with machine learning algorithms. 
This shows that the best performing models can achieve and unweighted average recall of .7.
\end{abstract}

\section{Introduction}\label{sec:intro}

A strong signal of emotional communication between humans are non-linguistic sounds such as laughter, coughs, or sighs. They can influence the semantics of utterances for example in the case of irony and generally add to the pragmatics of a communicative act by setting a context. 

As such, it is of great relevance to investigate and categorize non-linguistic sounds and  build machine learning models for them to be used for automated dialog enhancement.
This paper deals with a binary distinction of laughter with respect whether it is a signal of joyful togetherness or in contrary one of hurtful repudiation. Or to make it short: whether it is \textit{laughing with} or \textit{being laughed at}.
On this behalf, we conducted a data collection on the basis of Google AudioSet \cite{audioset} by selecting a subset of the samples annotated with \textit{laughter} based on being present in a communicative act with clear connotation of being either positive or negative.
On the basis of this annotated data, we performed two experiments: on the one hand, we manually extracted and analyzed phonetic features. On the other hand, we executed several machine learning experiments by systematically combining several automatically extracted acoustic feature sets with machine learning algorithms. 


Contributions of this paper are:
\begin{itemize}
    \item Compiling a database of annotated laughter.
    \item Distinguishing positive from negative laughter, which has to our best knowledge not been done before.
\end{itemize}

\section{Related Work}\label{sec:litref}

The acoustical analysis of laughter begins as early as 1899 with Boekes study on gelastic signal features \cite{Boeke1899}. In this early stage, the fundamental frequency, the number, and the duration of ha-sounds are determined using a phonogram.

In the late 20th century, empirical measurements are still done with a focus on the fundamental frequency and temporal features, like the length of impulses or the total laugh duration \cite{Mowrer1987} \cite{Rothgnger1998} with only some minor interest in the spectral features of laughter sounds \cite{inbook}.  

With the rising of new technological possibilities, basic parameters can be investigated more in detail. The vocalic nature of laughter is assessed thoroughly at the beginning of the 21st century \cite{Bachorowski2001} with a new interest for the link between acoustic cues and laughter evaluation \cite{Todt2001}. This sets a new pivot for further research, which rotates around the question how positively or negatively laughter variations are perceived \cite{Lavan2015}
\cite{Wood2020}.  One crucial finding is that variations of the acoustic properties of laughter are correlated to the laughter-eliciting situations \cite{Wood2017}.

In \cite{2013}, then, a modified variant of Wundt's model of emotional dimensions is implemented to investigate how arousal, dominance, sender's valence, and receiver-directed valence are linked to the acoustic properties of different discrete laughter categories (joyful laughter, taunting laughter, tickling laughter, and schadenfreude). According to the study's results, the laughter categorizations  can be reliably differentiated based on the different dimensional expressions, with a consistently better (88\,\%) recognition rate for positively perceived laughter types than for negatively perceived types (77\,\%) (Szameitat et al.\ 2013, p.\ 198). 

One of the most recent related projects is the MULAI database \cite{jansen-etal-2020-introducing}, a multimodal collection of laughter occurrences of about six hours that distinguishes mainly conversational laughter from joke induced ones and is annotated with the degree of humorousness.
Because it was collected with students, the age and social group is rather homogeneous. 

\section{The Laughter Corpus}\label{sec:data}

For the purpose of getting  preferably authentic stimuli, the database ``AudioSet'' \cite{audioset} 
from Google Research is used. Key advantages of this database are the usage of license-free YouTube videos, the large number of relevant laughter recordings, and the naturalness of the laughter contexts.

As to demographic criteria, only male laughter is included. The age limit is critical though. In this regard, there are no specifications or selection criteria to be found in the literature. So, if the subjects were perceptively still in the puberty vocal change or looked younger than 21 years old, they were excluded.
Another crucial criterion is the context. Only when sufficient context has been conveyed in the video to reliably assess the spontaneity and authenticity of the laughter episode, inclusion has been evaluated. The context really is particularly important, as it also ensures the categorical identification of the laughter episodes. 
The final selection criterion is audio quality. Excluded were any recordings that had a high level of noise, overload, or overlapping of the laughter signal with other noise sources. 
The selected 90 videos (45 for each valence-based laughter category) were downloaded and labeled according to the questionnaire, with ``a'' for happy laughter and ``b'' for mocking laughter, plus a unique numerical identifier. Subsequently,  the videos were converted to WAV-format, the individual laughter episodes extracted in PRAAT \cite{praat}, and the intensities normalized to a general maximum amplitude of 70\,dB. 

For reproducibility, the resulting database can be downloaded from a Zenodo repository\footnote{\url{https://zenodo.org/record/7224784}}.

\section{Perception Experiment}\label{sec:perc_exp}

For the perception experiment, two online applications were used: ``lab.js''\footnote{\url{https://lab.js.org/}} and  ``Open Lab''\footnote{\url{https://open-lab.online/}}, a free online study builder, which includes a randomization option for the stimuli presentation. By answering the questions about the auditory impression, the recognition rates of the laughter types are recorded. Since the stimuli were selected on the basis of the context, the ground truth is pre-determined. For each stimulus it can thus be defined, which response options are correct. 


\begin{figure}[h!]
  \centering
  \includegraphics[width=8cm]{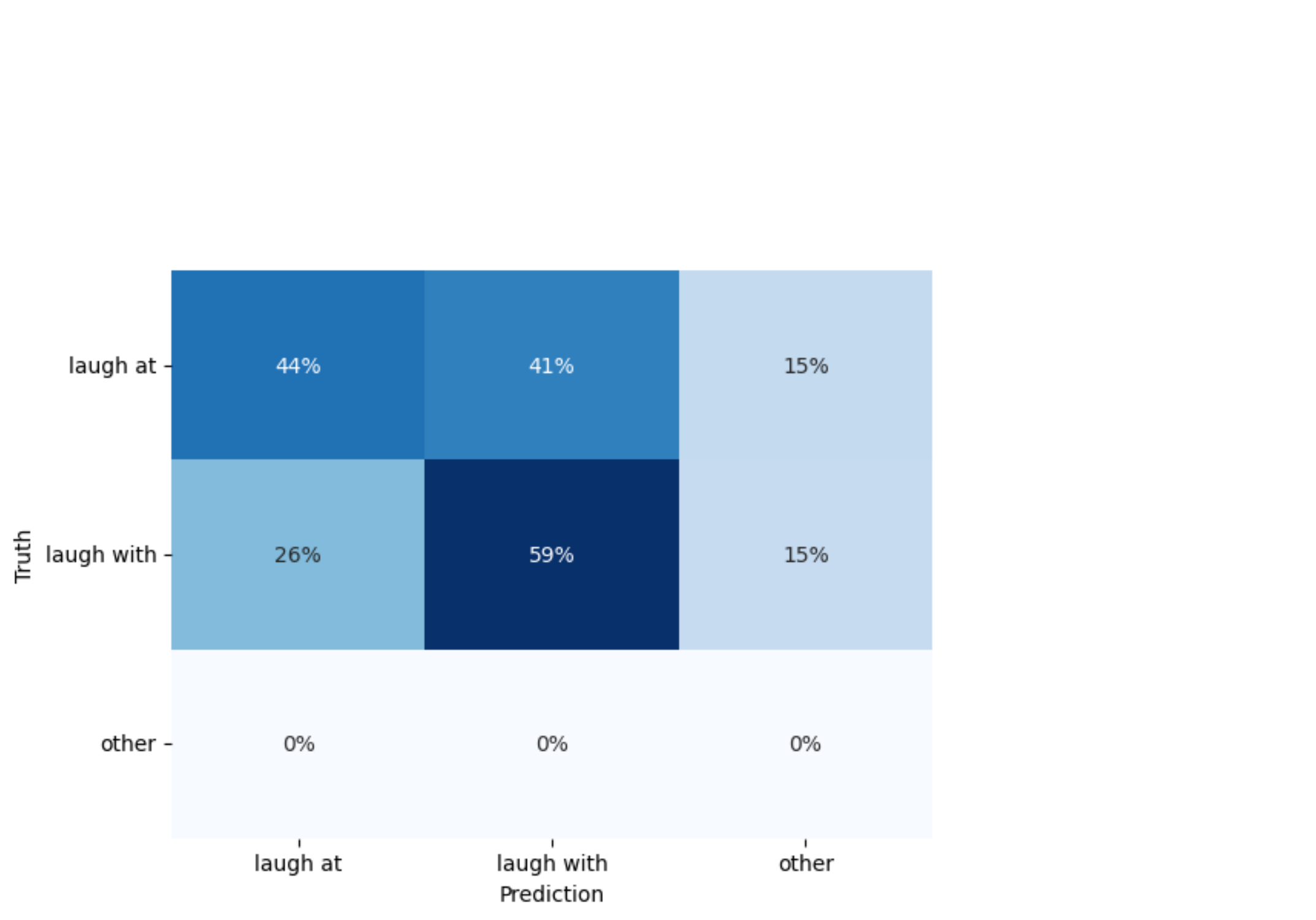}
  \caption{Confusion matrix for the laughter types (truth) and the listener judgements (predicted). \textit{other} is a label option not present in the data.}
  \label{fig:listener_conf}
\end{figure}

Altogether, nine female and eleven male participants between the ages of
23 and 35\,years were interviewed. The access link was sent to all participants via email, so that they could participate from any given location. A special focus was put on the instruction to consider oneself as the recipient of the laughter. That is, not to evaluate the laughter in the mindset of an external observer, but a direct participant in the interaction. 

For the experiment itself, participants had to answer the question about how the laughter stimulus is most likely felt, by selecting one of the following three answer categories: 
\begin{itemize}
    \item \textbf{a)} pleasant, friendly and / or affectionate
    \item \textbf{b)} unpleasant, aggressive and / or hostile
    \item \textbf{c)} other 
\end{itemize}

In Figure \ref{fig:listener_conf}, we show the confusion matrix for the listener judgements. It shows that both laughter types were almost equally often labeled as \textit{other}, but while the \textit{laugh at} type was not really distinguished, the \textit{laugh with} type shows a percentage of recognition strongly above chance level.

\section{Manually Extracted Acoustic Features}\label{sec:manual_feats}

A total of 19 relevant parameters have been derived from the cited literature. All of them proved to have a correlation with the perception of valence differences.

For the description of the spectral characteristics, the F0-minimum and -maximum, the mean absolute slope, F0-range and -standard deviation, F0- as well as F1- and F2-mean values, the center of gravity, and finally the peak frequency were measured.
Voicedness is assessed based on the percentages of voiced and unvoiced frames of the signal as well as on the Harmonics-to-Noise Ratio (HNR).
To determine the intensity, the minimum, maximum, mean, and standard deviation as well as the range of the sound pressure levels were measured. 
Finally, for the description of the temporal features, the total duration as well as the laughter rate (impulse per second) are measured.
The analysis of the stimuli expressions is done manually in Praat.

\section{Machine learning experiments}\label{sec:ml}
In order to further investigate the acoustic difference between the two kinds of laughter, we conducted a series of machine learning experiments.
For the experiments, we employed the Nkululeko framework\footnote{\url{https://github.com/felixbur/nkululeko/}} \cite{nkululeko} with an XGBoost classifier\footnote{Here, we use the Python XGBoost package: \url{https://xgboost.readthedocs.io/}.} \cite{Chen:2016} with the default meta parameters ($eta = 0.3$, $max\_depth = 6$, $subsample = 1$). This classifier is basically a very sophisticated algorithm based on classification trees and has been working quite well in many of our experiments \cite{burkhardt_2021}.

Additionally we use Support Vector Machines \cite{svm} -- a very well known algorithm discriminative learning optimally placing a separating hyperplane in a transformed feature space.
We use the Scikit-learn implementation with the standard C-value of $1.0$ and linear kernel. 
We consider the following acoustic features:
\begin{itemize}
    \item \textbf{os}: openSMILE \cite{opensmile} feature sets. They are based on frame-based low level descriptors such as $F_0$ combined by statistical functionals. The open source Python version of openSMILE\footnote{ \url{https://github.com/audeering/opensmile-python}} is being used and all  available features sets can be specified. 
    The default is the eGeMAPS set \cite{egemaps}, an expert set of 88 acoustic features. These features are being used in numerous articles in the literature as baseline features  \cite{Ringeval2018,compare16,burkhardt_2021}, as they work reasonably well with many tasks and are easy to handle for most classifiers based on their small number.  
    \item \textbf{mld}: Mid level descriptors as described in \cite{reichel2020}. Based on openSMILE low level descriptors, a syllabification is performed, and features describe suprasegmentals.
    \item \textbf{xbow}: openXBOW\footnote{\url{https://github.com/openXBOW/openXBOW}} -- the Passau Open-Source Crossmodal Bag-of-Words Toolkit \cite{xbow},
    an approach to use Bag-of-words techniques, known from natural language processing (NLP) that counts basic elements, for audio processing. The audio words are based in this case again on openSMILE low level descriptors. 
    \item \textbf{trill} TRILL embeddings \cite{trill}, meaning the penultimate layer of a deep neural network that has been trained with a triplet loss that distinguishes frames from the same speech file by those from different ones. TRILL has been trained on the speech tagged samples of the Audio Set database \cite{audioset}, which is about 2,793\,hours of audio data and has been published by Google and been evaluated on several domains, including emotional speech.
    \item \textbf{wav2vec}: Wav2Vec 2.0 \cite{wav2vec} embeddings also use the penultimate layer of a pre-trained deep neural network. This one has been trained by an approach to adopt the idea of the (again) NLP domain of word embeddings that model semantics by aligning words that occur in similar environments, to the audio domain. Although Wav2Vec primarily targets speech recognition (ASR), it is well suited to model speaker characteristics as well. There are several pre-trained versions that have been published by Facebook.
\end{itemize}

\begin{table}[h!t]
\centering
\begin{tabular}{l|rr}
feats/classifier & XGB  & SVM  \\\hline
Manual features (Sec. \ref{sec:manual_feats})  & .689 & .644 \\
OS eGeMAPS       & .511 & \bf{.700} \\
OS Compare\_16   & .566 & .611 \\
XBOW             & .566 & .588 \\
XBOW  with OS    & .488 & .666 \\
MLD              & .484 & .522 \\
MLD with OS      & .488 & .588 \\
TRILL emb.       & .577 & .666 \\
Wav2Vec emb.     & .577 & .611
\end{tabular}
\caption{Machine learning experimental results in UAR (unweighted average recall). XGB: XG-Boost, SVM: Support Vector Machine, OS: opensmile, XBOW: open crossbow, MLD: Mid Level Descriptors, TRILL: TRIpLet Loss network embeddings}
\label{tab:ml_results}
\end{table}

\begin{figure}[h!]
  \centering
  \includegraphics[width=8cm]{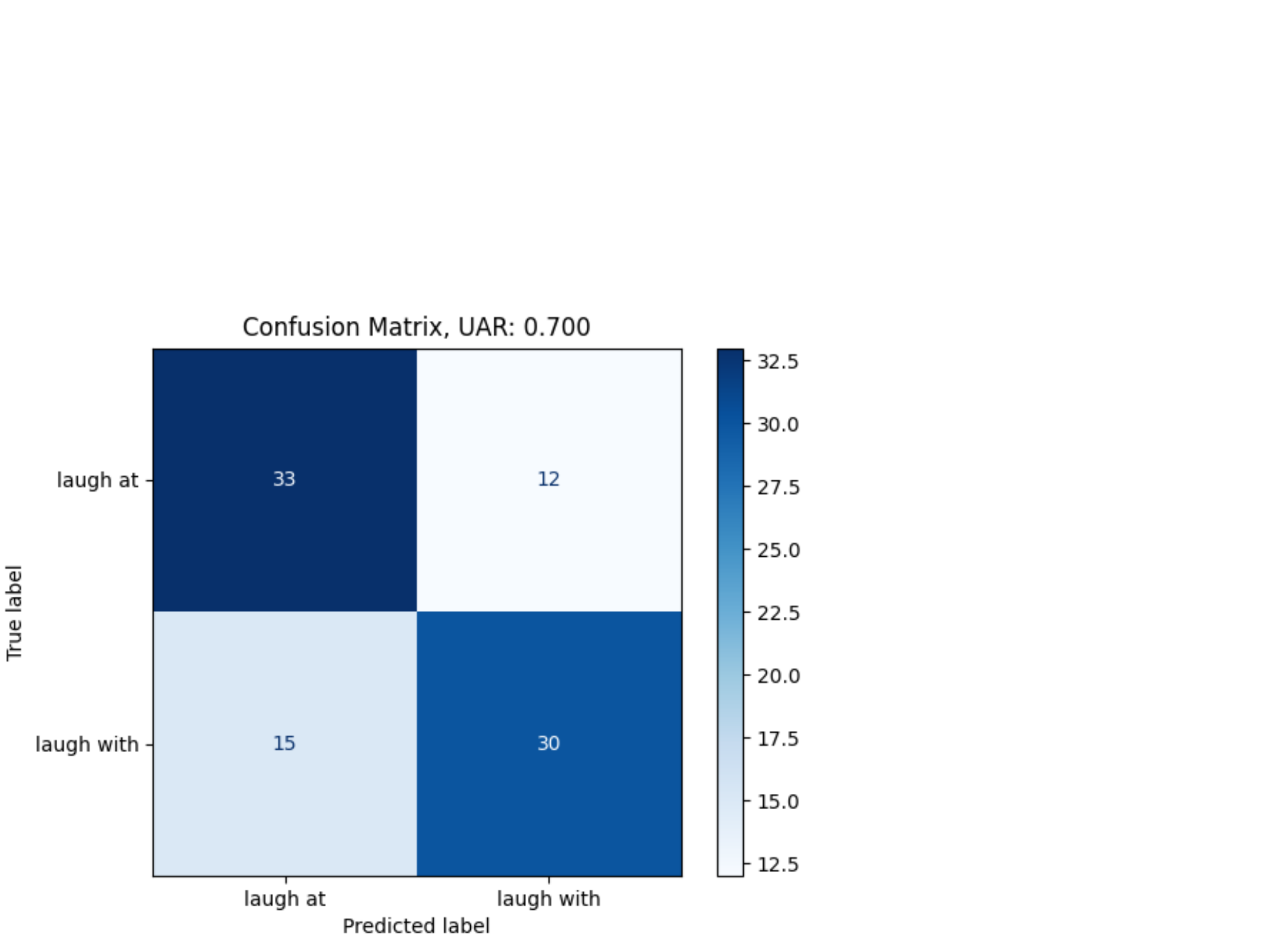}
  \caption{Confusion matrix for the SVM classifier with the openSMILE eGeMAPS features}
  \label{fig:confmat}
\end{figure}

We combined two machine classifiers with eight different sets of acoustic features, some of them expert-based (eGeMAPS, mid level descriptors and the ones described in Section \ref{sec:manual_feats}) and some of them learned (TRILL and Wav2Vec-2  embeddings).
Because there were only 90 samples, we did not split the data into training, development and test set, but did a 10-speaker group cross validation, meaning that we divided the speakers into 10 groups and then used them turn-wise as test set.
For reproducibility, the source code for the experiments is published on GitHub\footnote{\url{https://github.com/felixbur/happy_and_evil_laughter_data}}

\section{Discussion}\label{sec.discussion}

The main scope of this work was to analyse whether positive and negative laughter types can be 
automatically 
differentiated, and if so, which features separate taunting from joyful laughter. To answer this question, first of all, the perceptual experiment is indicative.

In the majority of cases (85.06\,\% ACC), 
either a negative or positive dimensioned nuance was perceived in the laughter signal. Total correct recognition is with 51.56\,\% (\textit{n} = 928)  well over the chance level. 
and similar to values obtained in the literature. As to joyful laughter, 87\,\% 
of the stimuli had been recognised correctly, compared to 64\,\% for mocking laughter. In summary, it can be stated that the investigated types of laughter differ in most cases on the basis of pleasant or unpleasant emotional feedback in the receiver. 
In a second step, it is analyzed which signal characteristics of the laughter types are crucial for the perception of different valences. For that purpose, a signal analysis is performed and the respective results are tested for significance. The results for the negative laughter type are as follows.

First, temporal features have been measured, i.\,e., the total duration and the laughter rate. Although there seems to be a tendency for shorter temporal extensions for both parameters, the difference between the two laughter types is not significant according to a T-test. 
The same is valid for the voicedness. Even though the data seems to suggest more harmonic energy, differences in Harmonic-to-Noise ratio (HNR) values
and the amount of voiced frames are not distinctive for any type.

Contrary, spectral measurements seem to be crucial. More specifically, the center of gravity, all F0-related measurements, F1 values, and peak frequencies are significantly (T-test, $p<.05$) higher for the stimuli of taunting laughter. That is, a higher pitch is perceived as more unpleasant. Therefore, an acute sound can be considered a marker for negative valenced laughter.

As to the F0 variability, smaller range, standard deviation,and slope have been observed. However, none of these seem to be a type-specific correlation.
Finally, as to intensity parameters, the significantly higher mean intensity and the, on average, lower intensity maxima outline a somewhat louder, but also more monotonous, laughter plateau with smaller maximum excursions. The values for the standard deviation and the range also indicate a lower intensity variability, although significance is not given.

The results of the machine learning experiment is shown in Table \ref{tab:ml_results}, the evaluation measure being unweighted average recall which is in this case identical to accuracy as the classes are evenly distributed. 
Not all features sets resulted to above chance classification. We do not see a pattern whether expert features (openSMILE eGeMAPS, mid level descriptors, Section \ref{sec:manual_feats}) work better or worse than learned features (TRILL and Wav2Vec-2 embeddings).
The best result was achieved with a SVM and openSMILE eGeMAPS features, the confusion matrix can be seen in Figure \ref{fig:confmat}.



\section{Conclusion}\label{sec:summary}

In summary, type-specific parameter expressions of laughing at and laughing with can be attested. Here, at least the significant variations of the spectral as well as some intensity features can be assumed to be idiosyncratic acoustic correlates of contrasting valence perceptions.
A machine learning experiment could support the assumption that indeed two different kinds of laughter were identified. The manually extracted features performed like other expert features.
%
Future work should revisit the topic on larger sets of laughter.

\section{Acknowledgements}
This research has been partly funded by the European EASIER (Intelligent Automatic Sign Language Translation) project (Grant Agreement number: 101016982). 

\bibliographystyle{IEEEtran}
\bibliography{literature}
\end{document}